\documentclass[article]{aa}

\makeatletter
\let\AA@orig@enddocument\enddocument
\AtBeginDocument{\let\enddocument\AA@orig@enddocument}
\makeatother

\renewcommand\makeLineNumber{}

\usepackage{graphicx}    
\usepackage[varg]{txfonts}             
\usepackage[dvipsnames]{xcolor}
\usepackage{hyperref}
\hypersetup{pdfborder = {0 0 0},
    colorlinks = true,
    linkbordercolor = {white},
    citecolor=NavyBlue,
    urlcolor=Green
}

\newcommand{\orcidlink}[1]{\protect\href{https://orcid.org/#1}{\protect\includegraphics[width=8pt]{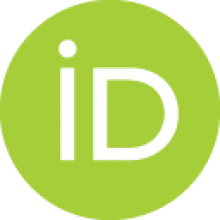}}}

\usepackage{amssymb}
\usepackage{amsmath}
\usepackage{xspace}
\usepackage{array,multirow}
\usepackage{enumitem}
\usepackage[english]{babel}
\usepackage{mathrsfs}
\usepackage{listings}
\usepackage{blindtext}
\usepackage{lipsum} 
\usepackage{booktabs}
\usepackage{url}
\usepackage{natbib,twoopt}
\usepackage[normalem]{ulem}  
\usepackage{placeins}

\newcommand{\ie}{i.e.\@\xspace} 
\newcommand{\eg}{e.g.\@\xspace} 

\renewcommand{\eqref}[1]{Eq.~\ref{#1}}
\newcommand{\fref}[1]{Fig.~\ref{#1}}

\newcommand{\tref}[1]{Table~\ref{#1}}
\newcommand{\sref}[1]{Sect.~\ref{#1}}

\newcommand{\aref}[1]{Appendix~\ref{#1}}

\newcommand{\kp}{\emph{Kepler}\xspace}

\newcommand{\teff}{\ensuremath{T_{\rm eff}}\xspace}

\numberwithin{equation}{section}

\makeatletter
\def\maketag@@@#1{\hbox{\m@th\normalfont\normalsize#1}}
\makeatother

\bibpunct{(}{)}{;}{a}{}{,} 

\makeatletter
\newcommand*\mysize{%
  \@setfontsize\mysize{5.7}{8.0}%
}
\makeatother

\makeatletter
\newcommand*\tabsize{%
  \@setfontsize\tabsize{7.}{8.0}%
}
\makeatother

\makeatletter
\newcommand\footnoteref[1]{\protected@xdef\@thefnmark{\ref{#1}}\@footnotemark}
\makeatother

\begin{document}

   \title{Bolometric corrections of stellar oscillation mode amplitudes\\ as observed by the PLATO mission}

   \subtitle{I. Planck-spectrum estimates}
    \titlerunning{PLATO bolometric corrections}

   \author{Mikkel N. Lund\inst{\ref{DK}}\orcidlink{0000-0001-9214-5642} \and J\'{e}r\^{o}me Ballot\inst{\ref{JB}}\orcidlink{0000-0002-9649-1013} \and William J. Chaplin\inst{\ref{WJC}}\orcidlink{0000-0002-5714-8618}
    }
    \offprints{MNL, \email{mikkelnl@phys.au.dk}}

    \institute{
    Stellar Astrophysics Centre, Department of Physics and Astronomy, Aarhus University, Ny Munkegade 120, DK-8000 Aarhus C, Denmark\label{DK} \and Institut de Recherche en Astrophysique et Planétologie (IRAP), Université de Toulouse, CNRS, CNES, 14 avenue Edouard Belin, 31400 Toulouse, France\label{JB} \and School of Physics and Astronomy, University of Birmingham, Edgbaston, Birmingham B15 2TT, UK\label{WJC} }

\authorrunning{Lund et al.}
   \date{Received December 18, 2025; accepted March 9, 2026}

 
  \abstract
   {}
   {We derive bolometric correction functions for oscillation mode amplitudes observed by the different cameras of the ESA PLATO mission. Such corrections between bolometric (full light) and mission instrument-specific amplitudes enable comparisons to theoretical expectations and amplitude conversion between different photometric missions, which is essential for proper detectability yields and target selection.}
   {Bolometric correction functions were calculated assuming a Planck function approximation for the stellar spectral flux distribution. The calculations follow the procedures applied in earlier analyses for the NASA \kp and TESS missions. We derived power-law and polynomial parametrisations of the bolometric corrections with \teff.}
   {We find that on average, oscillation mode amplitudes from PLATO's normal cameras (N-CAMs) are expected to be ${\sim}6.7\%$ lower compared to \kp, and ${\sim}12.5\%$ higher compared to TESS. A significant average amplitude ratio of ${\sim}25\%$ is expected for amplitudes measured using the blue PLATO fast camera (F-CAM) compared to TESS. We find that observations of bright solar-like oscillators, especially with PLATO's F-CAMs, would provide an important test of the predicted corrections.}
   {}

   \keywords{Asteroseismology --
                methods: data analysis -- stars: oscillations (including pulsations) -- stars: solar-type
               }

   \maketitle

\section{Introduction}

The ability to predict the amplitudes of solar-like oscillation modes for a space-based photometric mission, based on theoretical or empirical expectations, is essential to enable informed target selection and for realistically estimating asteroseismic detectability yields \citep{Chaplin2011,Schofield2019,Goupil2024}. Furthermore, the ability to convert observed mode or granulation amplitudes obtained from a given mission \citep[\eg][]{Huber2011,Corsaro2013,Sayeed2025} to their bolometric counterparts is required to, for example, appraise the quality of theoretical models for mode excitation \citep{Houdek1999,Zhou2019,Zhou2021}.

We provide an estimate of the bolometric correction $c_{\rm BP-bol}$ for the ESA PLATO mission \citep{Rauer2025}, scheduled for launch in late 2026. \sref{sec:bp} describes the PLATO spectral response functions used in the calculations of $c_{\rm BP-bol}$, while \sref{sec:res} provides the results for PLATO's `Normal' (N-CAM) and `Fast' (F-CAM) camera $c_{\rm BP-bol}$ values based on Planck spectra, including comparisons to the CoRoT, \kp, and TESS  photometric missions \citep{M09, B11, Lund2019}. 
In \sref{sec:dis} we discuss the impact and use of the $c_{\rm BP-bol}$ values.
\section{PLATO spectral response functions}\label{sec:bp}

An essential ingredient for computing the mode amplitude bolometric correction factor is the spectral response function ($\mathcal{S}_{\lambda}$) of the instrument---providing, as a function of wavelength, the fraction of incident light that contributes to a measured photometric signal. While the absolute values of this throughput are important for the photometric noise level and flux-to-magnitude conversion \citep{Marchiori2019,Borner2024,Jannsen2024}, here we are only sensitive to the relative values of $\mathcal{S}_{\lambda}$ with wavelength. For the following description, we refer, in particular, to the `PLATO Instrument Performance Report' \citep[PIPR;][]{PIPR}.
\begin{figure*}
   \centering
   \includegraphics[width=\textwidth]{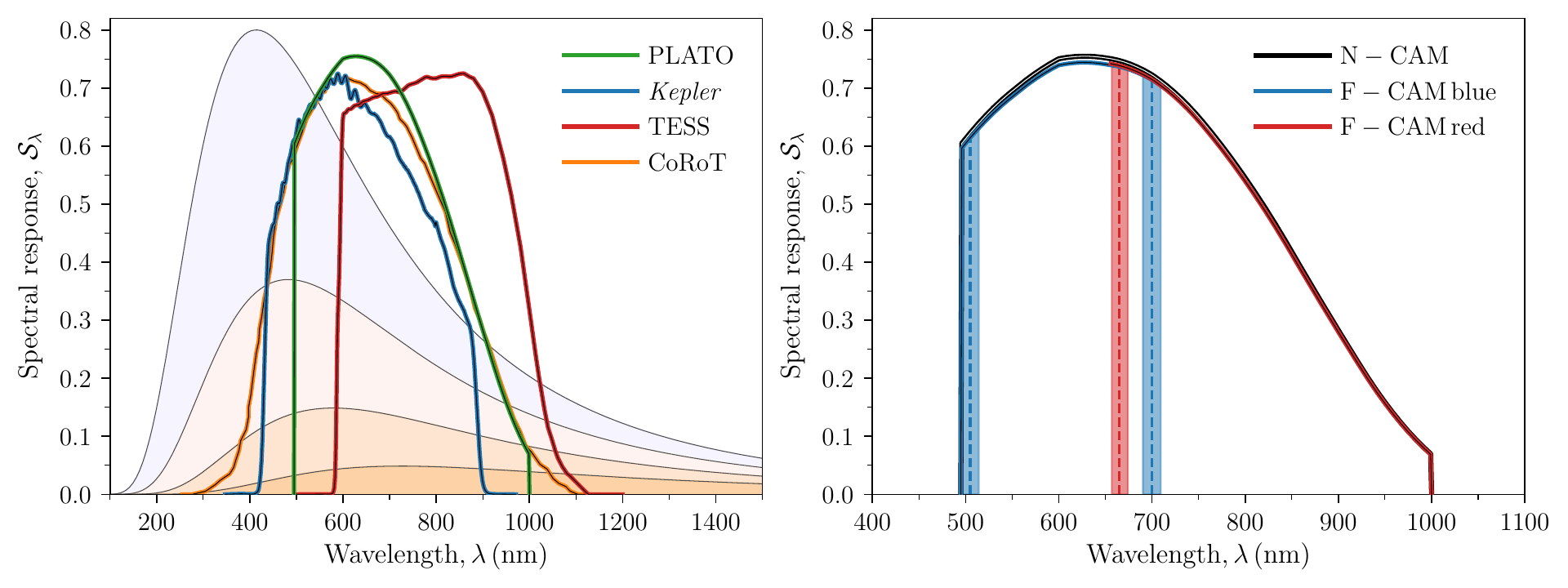}
      \caption{Left: Spectral response functions, $\mathcal{S}_{\lambda}$, for the PLATO (\sref{sec:bp}), \kp \citep{VCleve}, TESS \citep{2014SPIE.9143E..20R} (normalised to a maximum of \kp), and CoRoT \citep{2009A&A...506..411A} missions as a function of wavelength $\lambda$. The shaded regions show blackbody spectra with temperatures of $7000, 6000, 5000$, and $4000$ K (normalised to a maximum of $0.8$ for the hottest, or bluest, curve). Right: Spectral response functions for the PLATO N- and F-CAMs. The vertical dashed lines show the nominal passband limits of the blue and red F-CAMs, with the shaded $\pm10$ nm regions indicating the potential variation in these from the incident angle. }
         \label{fig:bands}
\end{figure*}

\begin{figure*}
   \centering
   \includegraphics[width=\textwidth]{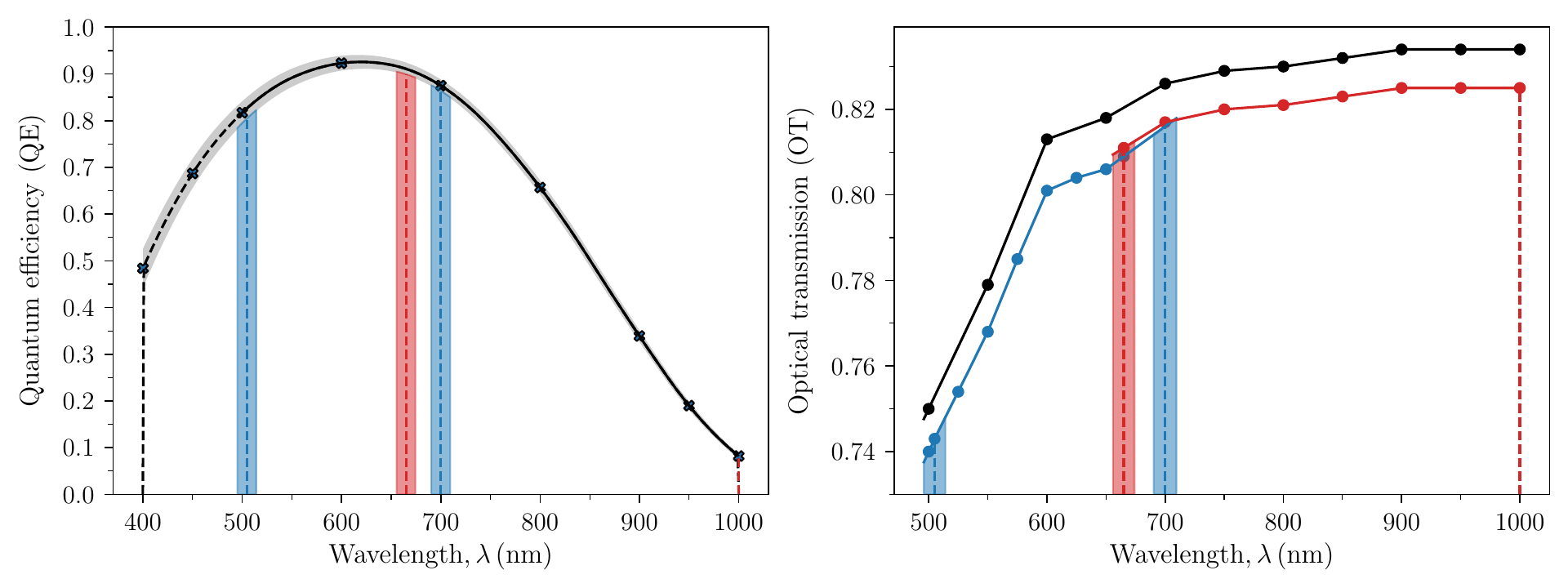}
      \caption{Left: QE as a function of wavelength used for computing $\mathcal{S}_{\lambda}$ (given by the product of QE and the OT in the right panel). The markers indicate the available averages from tests of N-CAM flight model CCDs, while the shaded region indicates the associated standard deviation. The full or dashed black line shows the adopted interpolation, where the dashed part indicates the wavelength range outside the PLATO passbands. The vertical dashed lines show the nominal passband limits of the blue and red F-CAMs, with the shaded $\pm10$ nm regions indicating the potential variation in these from the incident angle, with the upper red filter limit fixed at $1000$ nm by the detector response. Right: OT for the N-CAM (black) and F-CAMs (blue and red) as a function of wavelength. The markers indicate available measurement points. The F-CAM passband boundaries are marked as in the left panel. The full lines show a simple linear interpolation of the measurement points.}
         \label{fig:qeot}
\end{figure*}

For the N- and F-CAMs, we adopted for $\mathcal{S}_{\lambda}$ the so-called `current best estimate' (CBE; see PIPR table 3-1) of the beginning-of-life (BoL) mean combined photon conversion efficiency (PCE), as shown in \fref{fig:bands}. For the N-CAMs, the PCE used here is the product of the `as designed' CBE for on-axis optical transmission (OT; including worst-case levels of molecular and particulate contamination) and the mean flight model `as measured'\footnote{`Payload Straylight Analysis Report' (\texttt{PTO-EST-PL-REP-0936}) table 4, from `PLATO FM Deliverable Device EO Results' (\texttt{PTO-E2V-CCD-DP-1114}, issue 5.0)} charge-coupled device (CCD) quantum efficiency (QE); see \fref{fig:qeot}. The N-CAM passband covers the spectral range from $500$--$1000$ nm. 

For the F-CAMs, the OT is assumed to be the same as for the N-CAMs, as the optical design is identical; however, the PCE includes the additional impact of the red/blue spectral bandpass filters on the F-CAM window surfaces. These filters have a very high transmission\footnote{set to a constant value of 0.986 between 500 and 700 nm (blue) and 0.989 between 665 and 1100 nm (red).} (${\geq}98\%$) and hence mainly define the passband for the F-CAMS. The manufacturer limits specify boundaries of $505\pm10$ nm and $700\pm10$ nm for the blue filter, and a lower boundary of $665\pm10$ nm for the red filter, while the detector response defines an upper boundary at $1000$ nm. In this work, we assumed, as a baseline, the nominal values from these ranges. 
The QE is expected to remain unchanged during the mission, while the OT is expected to degrade slightly from the impact of radiation on glass transmission and coating efficiency, increasing charge transfer inefficiency \citep{Mishra2025}. Nevertheless, this effect is not expected to be wavelength-dependent. 

The uncertainty considered for $\mathcal{S}_{\lambda}$ comes from the uncertainty or dynamical range in the QE from tests of N-CAM full-frame CCD flight models. It lies at an ${\sim}3\%$ level at $500$ nm, and smaller for longer wavelengths. There is a small difference in the QE from the different CCDs used for the N-CAMs (full-frame CCD) and F-CAMs (frame-transfer CCD), but this is well within the included uncertainty.

In terms of potential systematic variations in $\mathcal{S}_{\lambda}$, the QE is known to be affected by temperature variations in a wavelength-dependent manner\footnote{See `Instrument Signal and Noise Budget' (\texttt{PLATO-DLR-PL-RP-0001}, issue 4.0)}. However, this is only at an ${\sim}1\%$ level for a 10K temperature range at $900$ nm, with nearly no variation below $700$ nm. Finally, we note potential impact from the incident angle: the QE increases by up to ${\sim}3-5\%$ in the wings of the spectral band, while decreasing by up to ${\sim}1-2\%$ at ${\sim}800$ nm for incident angles between $25$--$35^{\circ}$ \citep{Verhoeve2016}. Furthermore, the incident angle will affect the in-band to out-of-band transition wavelength of the F-CAM filters, shifting these to lower wavelengths at a level of $2-4$ nm at $10^{\circ}$ and $9-15$ nm at $20^{\circ}$ (PIPR).  

To construct the $\mathcal{S}_{\lambda}$ (\ie the product between QE and OT), we first interpolated the QE values using a second-order univariate spline, while the OT values for the N- and F-CAMs were linearly interpolated (more flexible interpolations generally led to unwanted variations near the F-CAM boundaries around the filter's overlap region). We note that, currently, no OT measurements exist at the $700$ nm boundary of the blue F-CAM; hence, beyond $665$ nm the $\mathcal{S}_{\lambda}$ for the blue F-CAM is given by linear extrapolation.  

While the fine details of the $\mathcal{S}_{\lambda}$ specifications might change before the launch of PLATO, the response adopted here should still provide a good baseline for understanding the bolometric correction. In what follows, we assess the impact of several potential small variations in $\mathcal{S}_{\lambda}$.

\section{PLATO bolometric corrections}\label{sec:res}

The bolometric correction $c_{\rm BP-bol}$ provides the conversion factor from oscillation mode (or granulation) amplitudes measured in the bandpass (BP) of a given photometric mission to bolometric amplitudes, \ie
\begin{equation}\label{eq1}
A_{\rm bol} = c_{\rm BP-bol}\, A_{\rm BP}\, ,
\end{equation}
and thereby different mission instrument-specific $c_{\rm BP-bol}$-values provide the conversion of amplitudes between missions:
\begin{equation}\label{eq:conversion}
A_{\rm BP_1} = (c_{\rm BP_2-bol}/c_{\rm BP_1-bol})\, A_{\rm BP_2}\, .
\end{equation}

In the current analysis, we consider only a blackbody (Planck) function to describe the stellar spectral flux density and refer to \citet{Lund2019} and \citet{B11} for details on the procedure for calculating $c_{\rm BP-bol}$. We consider the photometric missions of PLATO, TESS, \kp, and CoRoT. When using simply `PLATO', we refer to the values from the N-CAMs. Either `FCB' or `FCR' is added when referring to the blue or red PLATO F-CAMs.
\begin{figure}
   \centering
   \includegraphics[width=\columnwidth]{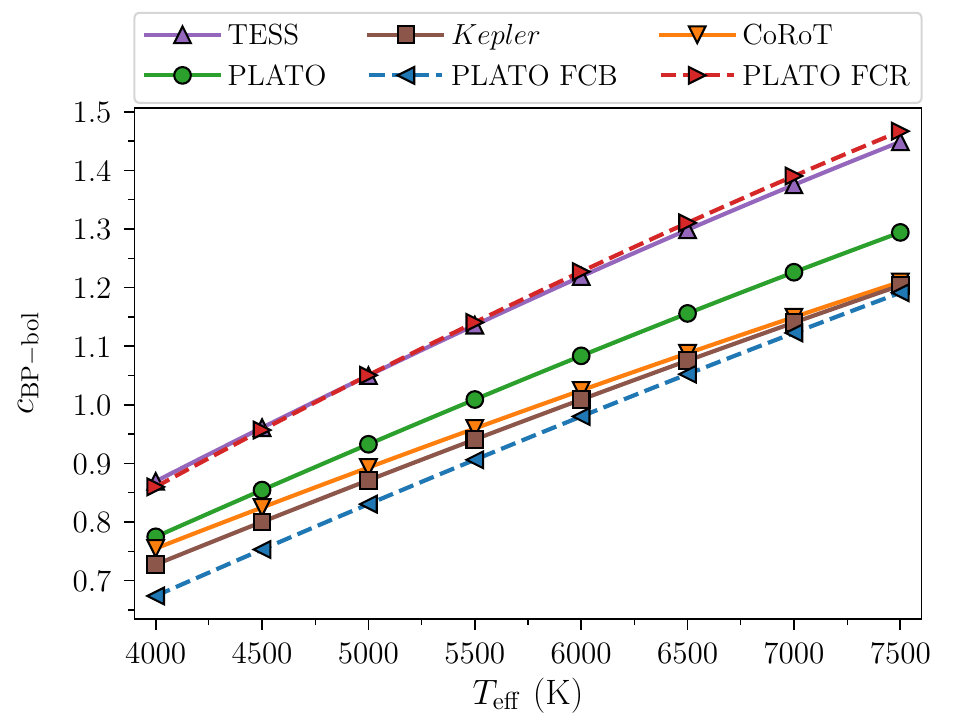}
      \caption{Values for the bolometric correction $c_{\rm BP-bol}$ (\eqref{eq1}) based on Planck spectra, as a function of \teff. Different lines and markers refer to different photometric missions or filters for the PLATO mission (see legend).}
         \label{fig:c_vs_T}
   \end{figure}

In \fref{fig:c_vs_T}, we show the calculated values of $c_{\rm BP-bol}$ for the different missions as a function of effective temperature and provide in \aref{sec:relations} (\tref{tab:rel_coeff}) the new power-law and polynomial relations for the PLATO mission, following the setup in \citet{Lund2019}. As expected from the PLATO N-CAM passband (\fref{fig:bands}), the $c_{\rm PLATO-bol}$ values lie between those of \kp and TESS, while these missions nearly overlap with, respectively, the blue and red F-CAM values.

\begin{figure}
   \centering
   \includegraphics[width=\columnwidth]{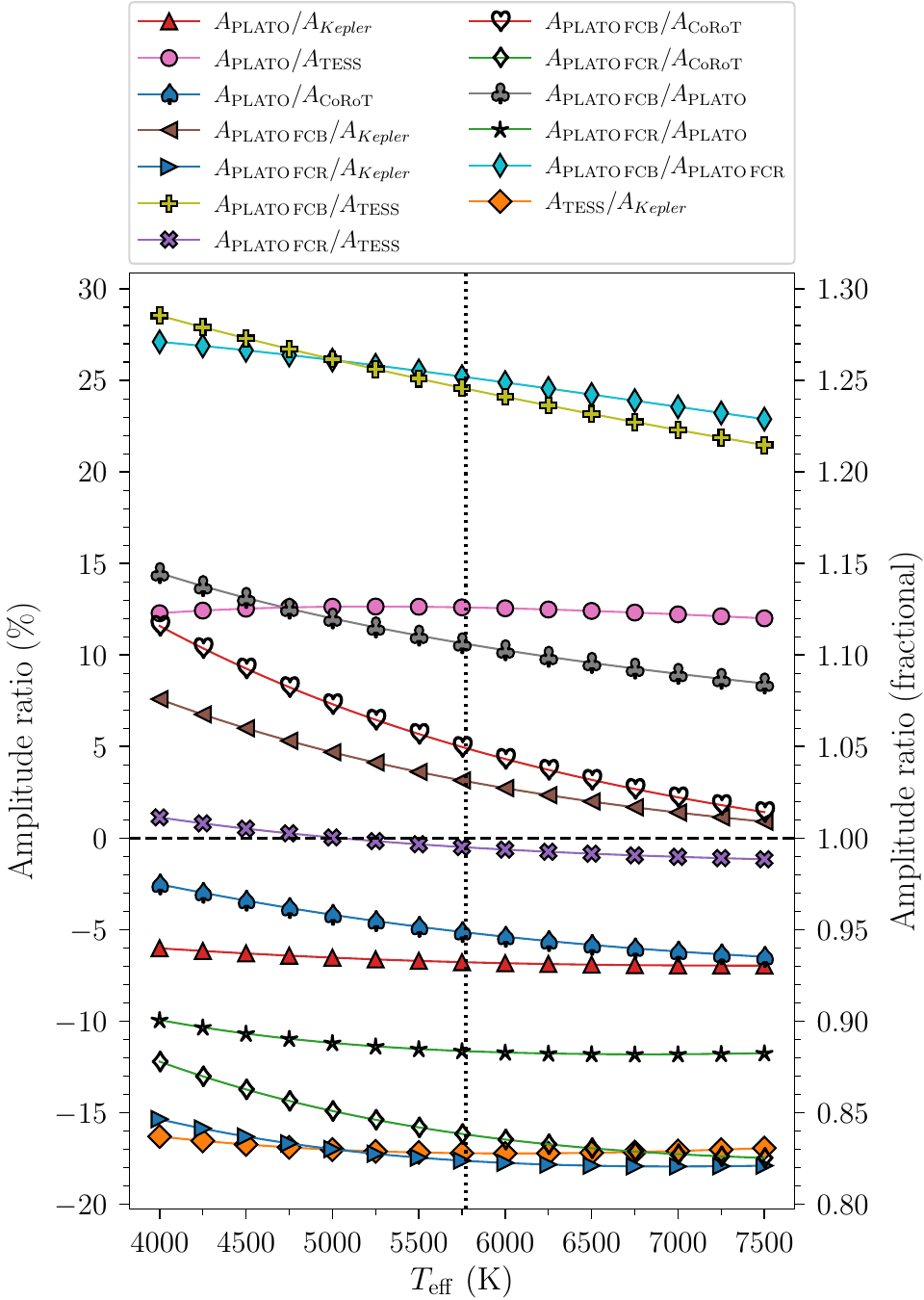}
      \caption{Amplitude ratios between different missions, and/or the different filters of the PLATO mission, as a function of \teff (see legend). The amplitude change is shown as both a percentage deviation (left axis) and as a fractional ratio (right axis). The horizontal dashed line indicates equal amplitudes, while the vertical dotted lines indicate the solar \teff.}
         \label{fig:c_ratios}
   \end{figure}

In \fref{fig:c_ratios}, we display several combinations of amplitude ratios (\eqref{eq:conversion}) between the missions and instruments. Across the \teff range considered, we find that on average oscillation mode amplitudes from PLATO (N-CAMs) can be expected to be ${\sim}6.7\%$ lower than \kp, and ${\sim}12.5\%$ higher than TESS. A significant amplitude difference is found between TESS and the blue PLATO F-CAM, with amplitudes being on average ${\sim}25\%$ higher for PLATO, peaking around ${\sim}29\%$ higher at the lowest \teff values. As expected from the overlap in $c_{\rm BP-bol}$ between TESS and the red PLATO F-CAM (\fref{fig:c_vs_T}), a similar amplitude ratio is expected between the blue and red F-CAMs.  

Considering the uncertainty in the spectral response via the QE (\sref{sec:bp}), we also estimated the impact on the $c_{\rm PLATO-bol}$ values of varying the QE linearly within a $\pm3\sigma$ band, starting from QE$\pm3\sigma$ at 500 nm and ending at QE$\mp 3 \sigma$ at 1000 nm. This resulted in only a ${\approx}\pm 0.15-0.20\%$ fractional change compared to the nominal value, with the version starting low at 500 nm (\ie the most redshifted) having the highest $c_{\rm BP-bol}$, corresponding to the lowest amplitudes. Similarly, we tested the expected change to the spectral response via the QE's sensitivity to the incident angle. This was done by changing the QE linearly from a $+5\%$ increase at 500 nm to a $-2\%$ decrease at 800 nm and then rising to a $+5\%$ increase at 1000 nm. This modification to the QE amounted to a mere $-0.22\%$ average change to $c_{\rm PLATO-bol}$. Overall, we consider these effects to have a negligible impact on $c_{\rm PLATO-bol}$.

A more significant change comes from considering the incident angle impact on the bolometric corrections for the PLATO F-CAMs. Here, in addition to the above modification to the QE, the boundaries of the F-CAM passbands are shifted blueward (\sref{sec:bp}). We assume a shift of the blue filter boundaries by $-10$ nm (\ie spanning the range $495-690$ nm) and similarly for the lower boundary of the red filter (now spanning $655-1000$ nm), corresponding to incident angles between $25$--$35^{\circ}$ \citep{Verhoeve2016}. As expected, the blueward shift or extension of the passbands decreases the $c_{\rm BP-bol}$ values for both filters, as they move closer to the peak of the Planck functions for the stars considered. The change is greater for the blue filter as both its boundaries shift, with an overall decrease in $c_{\rm BP-bol}$ by approximately $-1.5\%$. The red filter only extends slightly into the blue, resulting in a $-0.75\%$ decrease in $c_{\rm BP-bol}$. Following \eqref{eq:conversion} these decreases in $c_{\rm BP-bol}$ result in corresponding increases in the amplitudes from the F-CAMs compared to other missions. The amplitude ratio between the blue and red F-CAM filters ($A_{\rm PLATO\, FCB}/A_{\rm PLATO\, FCR}$) generally increases by $+1\%$ across the \teff range covered in \fref{fig:c_ratios}.
While several different combinations of perturbations could be imagined, the modifications considered above suggest that modest variations in the spectral response functions (N/F-CAMs) will generally not exceed the $\pm1-2\%$ level.

\section{Discussion}\label{sec:dis}
We provided relations for the bolometric conversion factors for the upcoming ESA PLATO mission, for both the normal (N) and the red/blue fast (F) cameras. This makes it possible to convert empirical amplitude relations from, for instance, the \kp mission to PLATO, which has a direct impact on the prediction of asteroseismic detectability yields. This should be considered in future updates of, say, the yields of \citet{Goupil2024}, where amplitude (and mode visibility) relations from \kp were adopted---which we can expect to be ${\sim}6.7\%$ higher than what PLATO's N-CAMs will observe.
We also note that simulation tools, such as PlatoSim \citep{Jannsen2024,Jannsen2025}, should include the provided bolometric corrections in their scaling of oscillation mode amplitudes, rather than, for example, the conversion by BP-integrated flux ratios currently included in PlatoSim.

With its F-CAMs, PLATO will provide an intriguing opportunity to directly test the $c_{\rm BP-bol}$ predictions \citep{Sreenivas2025}, at least in a relative sense. Mode amplitudes from the blue and red F-CAMs are expected to differ by up to ${\sim}25\%$ (\fref{fig:c_ratios}), so having simultaneous photometry from both F-CAMs for a given bright target would enable a direct test of the predictions. From TESS, we already know of several bright dwarf and subgiant solar-like oscillating stars positioned in the PLATO fields (\citealt{Lund2025}, Panetier et al., in prep.; see also \citealt{Eschen2024} and \citealt{Nascimbeni2025}), which could be good candidates for F-CAM observations. If the oscillations are detectable with TESS, they will certainly be detectable in both the PLATO N- and F-CAMs, yielding additional comparisons for the predictions.

In the current analysis, we approximated the stellar spectral flux densities by Planck spectra. In a follow-up analysis, we will, as in \citet{Lund2019}, appraise the impact of using synthetic spectra; include any potential updates to the PLATO response functions; and provide estimates for oscillation mode visibilities \citep{B11}.

\begin{acknowledgements}
We are grateful to Juan Cabrera, Sami Naimi, Demetrio Magrin, Francesco Borsa, Nicholas Jannsen and other members of the PLATO Performance Team for their assistance and discussions on the inputs to the PLATO spectral response functions used in this work.   
This work presents results from the European Space Agency (ESA) space mission PLATO. The PLATO payload, the PLATO Ground Segment and PLATO data processing are joint developments of ESA and the PLATO Mission Consortium (PMC). Funding for the PMC is provided at national levels, in particular by countries participating in the PLATO Multilateral Agreement (Austria, Belgium, Czech Republic, Denmark, France, Germany, Italy, Netherlands, Portugal, Spain, Sweden, Switzerland, Norway, and United Kingdom) and institutions from Brazil. Members of the PLATO Consortium can be found at \url{https://platomission.com/}. The ESA PLATO mission website is \url{https://www.cosmos.esa.int/plato}. We thank the teams working
for PLATO for all their work. MNL acknowledges support from the ESA PRODEX programme (PEA 4000142995). JB acknowledges the support from CNES, focused on PLATO. 
\end{acknowledgements}

\bibliographystyle{aa} 
\bibliography{MasterBIB} 

\begin{appendix}
\onecolumn
\section{Bolometric correction relations}\label{sec:relations}

\tref{tab:rel_coeff} provides the coefficients for the two parametrisations considered for the dependence of $c_{\rm BP-bol}$ on \teff. The subscript of the models refers to a functional form of either:
\begin{equation}\label{eq:rel1}
c_{P\rm -bol}(\teff) \approx \left(\frac{\teff}{T_o}\right)^{\alpha}\, ,
\end{equation}
for a subscript of `1', or a second-order polynomial:
\begin{equation}\label{eq:rel2}
c_{P\rm -bol}(\teff) \approx \sum_{i=0}^2 a_i(\teff-T_o)^{i}\, ,
\end{equation}
for a subscript of `2'. In addition to the subscript number, the adopted spectral response is also indicated (N-CAM, F-CAM blue (FCB), or F-CAM red (FCR)). As seen from the residual root-mean-square errors in \tref{tab:rel_coeff}, the polynomial function provides the closest match to the theoretical values (as expected from its additional parameters), and should therefore in general be adopted to approximate $c_{P\rm -bol}(\teff)$.

\begin{table*}[h!]
\centering 
\caption{Parameters of the bolometric correction relations.} 
\label{tab:rel_coeff}
\begin{tabular}{lcccccc} 
\cmidrule[1.0pt](lr){1-7}\\ [-1.8ex]
Model &$T_o$ & $\alpha$ & $a_0$ &  $a_1$ & $a_2$ & $\sigma_{\rm rms}$\\ 
  & $\rm (K)$ &  & $\rm (K)$ &  $\rm (K^{-1})$ & $\rm (K^{-2})$ & \\ 
\cmidrule(lr){1-7}\\ [-1ex]
$P_{\rm 1,\, N-CAM}$ & $5446$ & $0.81$		&  $\cdots$ 	& $\cdots$ 				& $\cdots$  	&		$1.73\times 10^{-3}$    \\
$P_{\rm 2,\, N-CAM}$ & $5446$ & $\cdots$	&  1 		& $1.512\times 10^{-4}$ 	& $-4.229\times 10^{-9}$ & $1.54\times 10^{-4}$ \\

$P_{1,\, \rm FCB}$ & $6137$ & $0.90$		&  $\cdots$ 	& $\cdots$ 				& $\cdots$  	&		$2.10\times 10^{-3}$    \\
$P_{2,\, \rm FCB}$ & $6137$ & $\cdots$  	&  1		&  $1.451\times 10^{-4}$ 	& $-3.530\times 10^{-9}$ & 		$1.90\times 10^{-4}$    \\

$P_{1,\, \rm FCR}$ & $4728$ & $0.84$		&  $\cdots$ 	& $\cdots$ 				& $\cdots$  	&		$4.19\times 10^{-3}$    \\
$P_{2,\, \rm FCR}$ & $4728$ & $\cdots$  	&  1		&  $1.874\times 10^{-4}$ 	& $-6.856\times 10^{-9}$ & 		$1.35\times 10^{-4}$    \\
\cmidrule[1.0pt](lr){1-7}\\ [-1ex]
\end{tabular} 
\tablefoot{The table provide the parameters for the power-law (\eqref{eq:rel1}; subscript `1') and polynomial (\eqref{eq:rel2}; subscript `2') model fits between \teff and $c_{\rm BP-bol}$ \citep[see][]{Lund2019}. The quality of the fits is quantified by the root-mean-square-error $\sigma_{\rm rms}$.}
\end{table*} 
\FloatBarrier

\end{appendix}
\end{document}